\documentclass[aps,prl,twocolumn,superscriptaddress,showpacs,floatfix,longbibliography,dvipsnames]{revtex4-2}
\pdfoutput=1
\usepackage{amsmath,amssymb,amsfonts,float,graphics,epsfig,epstopdf,xcolor,verbatim,tabularx,bm,multirow,appendix,hyperref}
\usepackage[normalem]{ulem}
\usepackage{lmodern}
\usepackage{ytableau}
\usepackage{pifont}
\usepackage{color}
\usepackage{bm}
\usepackage{wasysym}
\DeclareMathOperator{\Tr}{Tr}
\usepackage{ dsfont }

\hypersetup{
	colorlinks,
	linkcolor={BrickRed},
	citecolor={BrickRed},
	urlcolor={BrickRed},
	pdftitle={},
	pdfauthor={},
	pdfstartview={FitH}
}
\definecolor{myblue}{RGB}{65, 105, 225}

\begin{document}
	
\title{Origin of superconductivity in bilayer nickelates: a Quantum Monte Carlo study for a sign-problem-free effective model}

\author{Xu Zhang}
\affiliation{Department of Physics and Astronomy, Ghent University, Krijgslaan 299, 9000 Gent, Belgium}

\begin{abstract}
Using determinant Quantum Monte Carlo, we investigate the interplay between doping, inter-layer tunneling and onsite Hund's coupling in stabilizing superconductivity (SC) in a two-orbital model for the bilayer nickelate $\text{La}_3\text{Ni}_2\text{O}_7$.
With realistic dispersion and for certain values of the interaction parameters, the auxiliary-field-decoupled fermion Hamiltonian has Kramers anti-unitary symmetries which guarantee the absence of a sign problem. The same anti-unitary symmetries can also be used to show there is a second instability towards $(\pi,\pi)$ inter-layer bond order (BO) in the strong interaction limit. We indicate the possible connection between this bond order and the enigmatic density wave state observed in experiment, and clarify the decisive role played by the inter-layer tunneling in the competition between SC and BO. Finally, possible directions on how to enhance the SC transition temperature and stabilize the SC phase are also discussed.
\end{abstract}

\date{\today}
\maketitle

\noindent \textcolor{myblue}{\it{Introduction.-}}
In recent years, the discovery of high-$T_c$ superconductivity in bilayer nickelates $\text{La}_3\text{Ni}_2\text{O}_7$ under pressure~\cite{sun2023signatures,hou2023emergence,zhang2024high,wang2024pressure,zhang2024effects,wang2024bulk,li2025identification,zhou2025investigations,ueki2025phase,wang2023observation} has sparked an immense effort to understand the inner workings of these unconventional superconductors both experimentally~\cite{wang2024experimental,li2025ambient,khasanov2025pressure}
and theoretically~\cite{wang2025recent,lu2024interlayer,qu2024bilayer,chen2024orbital,chang2023fermi}. 
Density-functional theory (DFT) calculations~\cite{luo2023bilayer,zhang2023electronic,verraes2025first} have indicated the nearly quarter filled Ni-3$d_{x^2-y^2}$ and half filled Ni-3$d_{z^2}$ orbitals provide the dominant contributions to the bands near the Fermi energy. Pressure enhances the inter-layer tunneling between $d_{z^2}$ orbitals through oxygen $p$ orbital as well as an inter-layer antiferromagnetic super-exchange interaction $J_{zz}$, which together with the apical oxygen vacancies between two layers are believed to play important roles for superconductivity in the nickelates~\cite{dong2024visualization}. Nevertheless, the detailed pairing mechanism remains to be understood. In previous studies, some authors have stressed the necessity of hybridization between $d_{x^2-y^2}$ and $d_{z^2}$ orbitals~\cite{shen2023effective,yang2023interlayer,qin2023high,luo2024high,yang2024decomposition,lu2024interplay,ouyang2024absence,kaneko2024pair,tian2024correlation}, and others have pointed out the importance of the onsite Hund's coupling~\cite{lu2024interlayer,oh2023type,qu2024bilayer,zhang2024strong,pan2024effect,lange2024pairing,yang2024strong,tian2024correlation,lu2024interplay,ouyang2024absence,wu2024deconfined,oh2024high,kaneko2024pair,lange2024feshbach}. To understand how all these factors cooperate to give rise to the superconductivity observed in experiments, numerically-exact simulations serve as vital tools.

Among these numerical methods, density matrix renormalization group (DMRG) studies on quasi-one dimensional chains~\cite{qu2024bilayer} and ladders~\cite{kakoi2024pair,kaneko2024pair,lange2024feshbach,qu2025hund} can include both $d_{x^2-y^2}$ and $d_{z^2}$ orbitals in a Hamiltonian with realistic band dispersion, but it is impossible to fully extrapolate these results to the two-dimensional limit~\cite{schollwock2011density,cirac2021matrix}. On the other hand, a recent infinite projected entangled-pair state (iPEPS)~\cite{chen2024orbital} study could reach the correct thermodynamic limit, but was limited by the on-site Hilbert space dimension and could therefore not include both $d_{x^2-y^2}$ and $d_{z^2}$ orbitals~\cite{cirac2021matrix}. In contrast, determinant Quantum Monte Carlo (QMC), when the sign-problem is absent, is powerful in studying finite temperature strongly correlated fermion systems~\cite{assaad2008world,li2019sign,Xu2019revealing,pan2022sign}, e.g. in twisted bilayer graphene the QMC method has successfully confirmed the mean-field conjecture on the inter-valley coherent ground state at half filling~\cite{bultinck2020ground,pan2022dynamical,zhang2021momentum,hofmann2022fermionic,huang2025angle}. A single-orbital bilayer toy model with SU(2)$\times$SU(2) symmetry inspired by the nickelates has been studied with projective QMC~\cite{chang2023fermi}, where a 2+1D O(4) Wilson-Fisher quantum critical point was identified, but it is well-known that this high-symmetry model cannot support finite temperature order due to Mermin-Wagner theorem~\cite{mermin1966absence}. In this letter, inspired by the Hund’s-coupling-mediated superexchange pairing scenario~\cite{lu2024interlayer}, we point out the existence of a sign-problem-free limit of the two-orbital bilayer nickelate model, which allows for an efficient simulation and a study of the combined effects from the doping, inter-layer tunneling $t_{zz}^{\perp}$, and onsite Hund's coupling $J_{xz}$.

The paper is organized in three parts below. We first introduce the model and the set-up of the QMC simulations, identify candidate order parameters, and compare the single-particle dispersion between the free and interacting models. Then we study the role of the different parameters in the Hamiltonian in forming superconductivity (SC), and show the instability towards the competing inter-layer bond order (BO) and spin density wave (SDW) orders. Finally, we summarize our results and discuss the possible directions for improving SC critical temperature and further stabilizing the SC phase.

\noindent \textcolor{myblue}{\it{Hamiltonian.-}}
We use a Hamiltonian $H=H_0+H_I$ in our study, where parameters in $H_0$ are taken from the DFT calculations in Ref.~\cite{luo2023bilayer} and in $H_I$ only effects from onsite interactions are considered. The dominant inter-layer tunneling $t_{zz}^{\perp}$ combining with the onsite repulsion $U_{zz}$ of $d_{z^2}$ orbitals forms an effective antiferromagnetic interaction $J_{zz}\approx\frac{(t_{zz}^{\perp})^2}{U_{zz}}$ with the assumption $U_{zz}\gg t_{zz}^{\perp}$. The left parts related with $d_{z^2}$ orbitals are approximated as higher-order contributions. We ignore the onsite repulsion of $d_{x^2-y^2}$ orbitals, which can be justified self-consistently based on our low double-occupancy density result. For other interactions, we also take into account Hund's coupling $J_{xz}$ from exchange integral between $d_{x^2-y^2}$ and $d_{z^2}$ orbitals and ignore the inter-orbital repulsion and pair-hopping terms. The explicit Hamiltonian is
\begin{eqnarray}
	H_0&=&\sum_{\langle i,j \rangle} t_{mm}^{\parallel} c_{i,l,s,m}^\dagger c_{j,l,s,m} + \sum_{\langle\langle i,j \rangle\rangle} t_{mm}' c_{i,l,s,m}^\dagger c_{j,l,s,m} \nonumber\\
    &+&\sum_{i}t_{mm}^{\perp} c_{i,1,s,m}^\dagger c_{i,2,s,m} + h.c.+\sum_{i}\epsilon_m c_{i,l,s,m}^\dagger c_{i,l,s,m} \nonumber\\
    &+& \sum_{\langle i,j \rangle} t_{xz}^{\parallel}c_{i,l,s,x}^\dagger c_{j,l,s,z} + t_{xz}^{\parallel}c_{i,l,s,z}^\dagger c_{j,l,s,x} +h.c. \nonumber\\
    &+& \sum_{\langle i,j \rangle,l} t_{xz}'c_{i,l,s,x}^\dagger c_{j,\bar{l},s,z} + t_{xz}'c_{i,l,s,z}^\dagger c_{j,\bar{l},s,x} + h.c.,  \nonumber\\
    H_I&=&-J_{xz} \sum_{i}  \mathbf{S}_{i,l,x} \cdot \mathbf{S}_{i,l,z} + J_{zz} \sum_{i} \mathbf{S}_{i,1,z} \cdot \mathbf{S}_{i,2,z}.
\end{eqnarray}
Here we utilize $l\in\{1,2\},s\in\{\uparrow,\downarrow\},m\in\{x,z\}$ to label layer, spin and orbital ($d_{x^2-y^2}$ is labeled by $x$ and $d_{z^2}$ is labeled by $z$) degrees of freedom. Summation over repeated indices is implicit. $S^{\alpha}_{i,l,m}=\sum_{s,s'}c^{\dagger}_{i,l,s,m} \sigma^{\alpha}_{s,s'} c_{i,l,s',m}$ ($\alpha\in\{x,y,z\}$) are the spin components, $\langle i,j \rangle,\langle\langle i,j \rangle\rangle$ labels respectively the nearest and next-nearest neighbor hopping, and $\bar{l}$ is the opposite layer to $l$. Note the operators in inter-layer tunneling term $t_{zz}^{\perp}$ should be projected out the double-occupancy space. We cannot handle that in our QMC simulation but instead focusing two limits that the original $t_{zz}^{\perp}$ from DFT and $t_{zz}^{\perp}=0$. Working in units of eV below, the DFT parameters of $H_0$ are $t_{xx}^{\parallel}=-0.483,t_{zz}^{\parallel}=-0.110,t_{xx}'=0.069,t_{zz}'=-0.017,t_{xx}^{\perp}=0.005,t_{zz}^{\perp}=-0.635,\epsilon_x=0.776,\epsilon_z=0.409,t_{xz}^{\parallel}=\pm0.239,t_{xz}'=\mp0.034$~\cite{luo2023bilayer}, where $t_{xz}^{\parallel},t_{xz}'$ indicate the hopping with the sign structure fixed by the $d_{x^2-y^2}$ orbital symmetry along $x/y$ direction.

We now approximately rewrite the interaction in a form amenable to a Hubbard-Stratonovich (HS) transformation:
\begin{eqnarray}
    H_I&\approx& U\sum_{i} (\mathbf{S}_{i,1,x} - a \mathbf{S}_{i,1,z} + \mathbf{S}_{i,2,x} - a \mathbf{S}_{i,2,z})^2 \nonumber\\
    &-& U\sum_{i} (\mathbf{S}_{i,1,x} + a \mathbf{S}_{i,1,z} - \mathbf{S}_{i,2,x} - a \mathbf{S}_{i,2,z})^2 \nonumber\\
    &=&4U\sum_{i} \mathbf{S}_{i,1,x} \cdot \mathbf{S}_{i,2,x} - a \mathbf{S}_{i,1,x} \cdot \mathbf{S}_{i,1,z} \nonumber\\
    &-& a \mathbf{S}_{i,2,x} \cdot \mathbf{S}_{i,2,z} + a^2 \mathbf{S}_{i,1,z} \cdot \mathbf{S}_{i,2,z}
\end{eqnarray}
By taking $a=\frac{J_{zz}}{J_{xz}}$ and $U=\frac{J_{xz}^2}{4J_{zz}}$ one recovers the original interaction, but with an additional term $J_{xx}\sum_{i} \mathbf{S}_{i,1,x} \cdot \mathbf{S}_{i,2,x}$ where $J_{xx}=\frac{J_{xz}^2}{J_{zz}}$. Taking $J_{zz}\gg $$J_{xz}$ will make this additional term negligible. However, in this case we speculate the transition temperature is too low to detect numerically (the experimental transition temperature 80K$\sim0.007\,$eV~\cite{sun2023signatures,li2025ambient} corresponds to $\beta\sim143 \,\text{eV}^{-1}$ in contrast with the largest $\beta\sim24 \,\text{eV}^{-1}$ in our simulation). Thus when choosing parameters, we fix $J_{zz}=0.4$ (by taking $U_{zz}\approx1$) and tune $J_{xz}$ from $0.4$ to $0.1$ to study the Hund's coupling dependence.

To show that the modified interaction is sign-problem free, the partition function after Trotter decomposition and HS transformation is $Z=\Tr(\prod_{\tau}e^{-\Delta_{\tau} (H_0+H_I)}) \approx \sum_{\{g_{\tau,i,\alpha}\},\{g'_{\tau,i,\alpha}\}}\left[\prod_{\tau,i,\alpha}\frac{1}{16}\gamma(g_{\tau,i,\alpha})\gamma(g'_{\tau,i,\alpha})\right]\times\\\Tr\left[\prod_{\tau}\left( e^{-\Delta_{\tau}H_0}\prod_{i,\alpha}e^{i\sqrt{\Delta_{\tau}U}\eta(g_{\tau,i,\alpha})\hat{O}^{\alpha}_{i,1}}e^{\sqrt{\Delta_{\tau}U}\eta(g'_{\tau,i,\alpha})\hat{O}^{\alpha}_{i,2}}\right)\right]$. Here $\tau,i,\alpha\in\{x,y,z\}$ label imaginary time slice, lattice site and spin component and we pick $\Delta\tau=0.2$ in all simulations. $\gamma,\eta$ come from the discrete HS transformation $e^{\delta \hat{O}^2}=\frac{1}{4}\sum_{g=\pm1,\pm2}\gamma(g)e^{\sqrt{\delta}\eta(g)\hat{O}}+O(\delta^4)$, where $\gamma(\pm1)=1+\frac{\sqrt{6}}{3}, \gamma(\pm2)=1-\frac{\sqrt{6}}{3}, \eta(\pm1)=\pm\sqrt{2(3-\sqrt{6})}, \eta(\pm2)=\pm\sqrt{2(3+\sqrt{6})}$, and the definitions for $\hat{O}^{\alpha}_{i,1},\hat{O}^{\alpha}_{i,2}$ are $\hat{O}^{\alpha}_{i,1}=S^{\alpha}_{i,1,x} - a S^{\alpha}_{i,1,z} + S^{\alpha}_{i,2,x} - a S^{\alpha}_{i,2,z}, \hat{O}^{\alpha}_{i,2}=S^{\alpha}_{i,1,x} + a S^{\alpha}_{i,1,z} - S^{\alpha}_{i,2,x} - a S^{\alpha}_{i,2,z}$.

One can notice $-\Delta_{\tau}H_0$, $i\sqrt{\Delta_{\tau}U}\eta(g_{\tau,i,\alpha})\hat{O}^{\alpha}_{i,1}$, $\sqrt{\Delta_{\tau}U}\eta(g'_{\tau,i,\alpha})\hat{O}^{\alpha}_{i,2}$ are invariant under Kramers (i.e., anti-symmetric) anti-unitary transformation $T=l^x\sigma^ym^0K$, where $K$ is complex conjugation and $m^0$ is identity for orbital subspace ($c_{x},c_{z}$). According to Refs.~\cite{wu2005sufficient,li2015solving,li2016majorana,li2019sign,wei2016majorana}, this implies that the trace over fermionic degrees of freedom is positive definite for any auxiliary field configuration and hence gives rise to the well-defined probability weight for Monte Carlo sampling.

\noindent \textcolor{myblue}{\it{Anti-unitary symmetries and order parameters.-}}
Crucially, the same anti-unitary symmetries which guarantee the absence of the sign problem also imply that the SC pairing is maximal in the $\Delta^{\dagger}=c^{\dagger}l^x\sigma^ym^0c^{\dagger}$ channel~\cite{zhang2025constraints}. To analyze the competing orders of SC, we follow Ref.~\cite{zhang2025constraints} and first list all the Kramers anti-unitary symmetries of the interacting part. Afterwards, we will add back the kinetic terms and speculate the possible phase diagram. Working in the Majorana basis $\gamma_{1}=\frac{1}{\sqrt{2}}(c^{\dagger}+c), \gamma_{2}=\frac{i}{\sqrt{2}}(c^{\dagger}-c)$,
\begin{eqnarray}
    \hat{O}^{x/z}_{i,1}&=&\frac{1}{4}\gamma^{T}\kappa^{y}l^{0}\sigma^{x/z}((m^z+m^0)+a(m^z-m^0))\gamma, \nonumber\\
    \hat{O}^{y}_{i,1}&=&\frac{1}{4}\gamma^{T}\kappa^{0}l^{0}\sigma^{y}((m^z+m^0)+a(m^z-m^0))\gamma, \nonumber\\
    \hat{O}^{x/z}_{i,2}&=&\frac{1}{4}\gamma^{T}\kappa^{y}l^{z}\sigma^{x/z}((m^z+m^0)-a(m^z-m^0))\gamma, \nonumber\\
    \hat{O}^{y}_{i,2}&=&\frac{1}{4}\gamma^{T}\kappa^{0}l^{z}\sigma^{y}((m^z+m^0)-a(m^z-m^0))\gamma.
\end{eqnarray}
Here, $\kappa^\alpha$ are Pauli matrices acting in the 2-dimensional Majorana subspace ($\gamma_1,\gamma_2$).  It is easy to check $i\sqrt{\Delta_{\tau}U}\eta(g_{\tau,i,\alpha})\hat{O}^{\alpha}_{i,1}$, $\sqrt{\Delta_{\tau}U}\eta(g'_{\tau,i,\alpha})\hat{O}^{\alpha}_{i,2}$ are invariant under following Kramers anti-unitary symmetries
\begin{eqnarray}
    T_1&=&\kappa^{x/z} l^{x}\sigma^ym^{0/z}K, \\
    T_2&=&\kappa^{y} l^{x}\sigma^0m^{0/z}K, \nonumber\\
    T_3&=&\kappa^{0} l^{y}\sigma^0m^{0/z}K. \nonumber
\end{eqnarray}
Following Ref.~\cite{zhang2025constraints}, we now consider bilinear order parameters $\gamma^{T}(r) M^{i} \gamma(r)$ at space-time position $r$, where $M^i = \sigma^{\mu_1}\otimes\sigma^{\mu_2}\otimes...\otimes\sigma^{\mu_{n+1}}$ are $2^{n+1}$ dimensional matrices, with $\sigma^\mu = (\mathds{1},\boldsymbol{\sigma})$ and $\boldsymbol{\sigma}$ are three Pauli matrices. A fermion bilinear correlation function for a fixed auxiliary field configuration $\{g\}$ can be divided into two parts: a \emph{direct} part (the local contraction) and an \emph{exchange} part (non-local contraction):
\begin{eqnarray} \label{FW}
	&& \langle \gamma^{T}(r) M^{i} \gamma(r) \gamma^{T}(r') M^{i} \gamma(r')\rangle_{\{g\}}   \\
	& =& \langle \gamma^{T}(r) M^{i} \gamma(r) \rangle_{\{g\}} \langle \gamma^{T}(r') M^{i} \gamma(r')\rangle_{\{g\}} \nonumber\\
	& +&\frac{1}{2^{n}}
	\sum_{j}(-1)^{\eta_{ij}} \langle \gamma^{T}(r) M^{j} \gamma(r') \rangle_{\{g\}} \langle \gamma^{T}(r) M^{j*} \gamma(r') \rangle_{\{g\}}. \nonumber
\end{eqnarray}
In the last line we have used Fierz identity, $\eta_{ij}$ is defined via the relation $M^iM^j = (-1)^{\eta_{ij}} M^j M^i$, and the summation over $j$ runs over all different $M^j$. In Ref.~\cite{zhang2025constraints} it was proven that the exchange contribution to the correlation function is maximal for the spin-singlet layer-triplet SC $\overline{M}_{SC}=\langle c^{\dagger}l^x\sigma^ym^{0/z}c^{\dagger} \rangle$ because of the $T_1$ symmetry, as well as the inter-layer BO~\cite{chang2023fermi} $\overline{M}_{BO}=\langle c^{\dagger}l^{x/y}\sigma^0m^{0/z}c \rangle$ because of the $T_2,T_3$ symmetries. The direct contribution of SC/BO is zero due to charge conservation/layer charge conservation symmetry.

Let us now consider the effect of gradually adding back the kinetic terms $H_0$. First, the hopping $t_{xx}^{\parallel},t_{zz}^{\parallel},t_{xz}^{\parallel}$ with $\nu^x$ in sublattice space require modified particle-hole symmetries $\nu^zT_2,\nu^zT_3$, and also do not induce inter-layer tunneling such that the direct contribution remains zero for both SC and BO. $t_{xz}^{\parallel}$ picks up $m^0$ over $m^z$ for both SC and BO with its $m^x$ in orbital subspace and there is no difference between $\overline{M}_{SC}=\langle c^{\dagger}l^x\sigma^ym^0c^{\dagger} \rangle,\overline{M}_{BO}=\langle c^{\dagger}l^{x/y}\sigma^0m^0c \rangle$ except for $\overline{M}_{BO}$ structure factor peaking at $(\pi,\pi)$ due to the extra sublattice $\nu^z$ in the modified particle-hole symmetry.

The addition of terms $t_{xx}',t_{zz}',\epsilon_{x},\epsilon_{z}$ break particle-hole symmetries, but still preserve the layer charge conservation. As a result, the correlations of $M_{SC}$ are strictly larger than those of $M_{BO}$, which indicates that SC is preferred over inter-layer bond order. 

Finally, when adding back the inter-layer tunneling $t_{xx}^{\perp},t_{zz}^{\perp},t_{xz}'$, the layer charge conservation symmetry is broken and as a result the $M_{BO}$ correlation function will pick up a non-zero contribution from the direct contribution. The phase diagram is determined by the competition between the direct and exchange contributions. As a result, a transition from $\overline{M}_{SC}=\langle c^{\dagger}l^x\sigma^ym^{0}c^{\dagger} \rangle$ zero momentum order to $\overline{M}_{BO}=\langle c^{\dagger}l^{x}\sigma^0m^{0}c \rangle$ order at momentum $(\pi,\pi)$ is possible. We will see this borne out in our numerical simulations presented below.

\begin{figure}[htp!]
    \centering
    \includegraphics[width=\columnwidth]{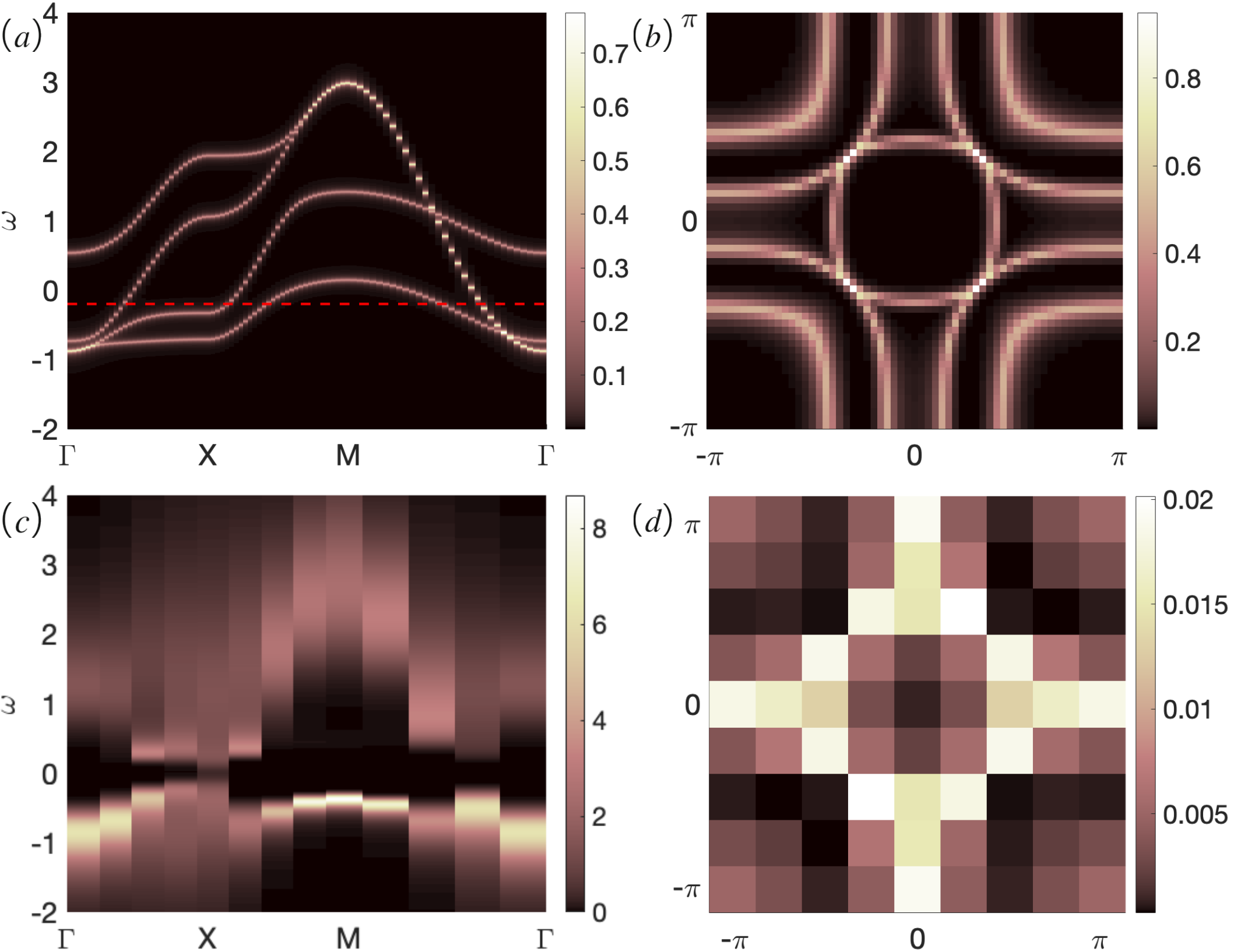}
    \caption{(a) The spectral function $A(k,\omega)$ for $H_0$ with parameters from DFT calculation~\cite{luo2023bilayer} with an artificial width $\delta=0.02$ to indicate the band structure along high symmetry line for size $L=60$, the dashed line indicates the chemical potential $\mu=-0.2$ used for interacting case (c)(d). (b) Plot of $\lim_{\beta \to \infty}G(k,\tau=\beta/2)\sim A(k,\omega=0)$ with $L=\beta=60$. The color scale indicates the zero frequency spectral weight, where the bright region indicates the position of Fermi surface. (c) Spectral function $A(k,\omega)$ along high symmetry line derived from stochastic analytical continuation (SAC)~\cite{shao2023progress} for interacting Hamiltonian with parameters $t_{zz}^{\perp}=-0.635,J_{xz}=0.4,L=8,T=0.05$. (d) The $G(k,\tau=\beta/2)$ with the same parameters as (c) showing the positions of single-particle excitation minima.}
    \label{fig:FSBD}
\end{figure}

\noindent \textcolor{myblue}{\it{Fermi surface and spectral function.-}}
The results for the single-particle excitations are shown in Fig.~\ref{fig:FSBD}. The chemical potential is tuned to $\mu=-0.2$ as the red line shown in Fig.~\ref{fig:FSBD}(a), where the doping hole density per site for $d_{x^2-y^2}$ and $d_{z^2}$ orbitals are converged to $\delta_{x^2-y^2}\approx0.47$ and $\delta_{z^2}\approx0.034$ at low temperature. The spectral function $A(k,\omega)$ for the non-interacting Hamiltonian $H_0$ with an artificial spectral width $\delta=0.02$ in Fig.~\ref{fig:FSBD}(a) shows the band structure along a high symmetry line. In Fig.~\ref{fig:FSBD}(b) we show the imaginary time Green's function $\lim_{\beta \to \infty} G(k,\tau=\beta/2)$ for the non-interacting model as a proxy for the zero frequency spectral function $A(k,\omega=0)$, which indicates the position of Fermi surface. For our QMC simulations of the interacting model, we take parameters $t_{zz}^{\perp}=-0.635,J_{xz}=0.4,L=8,T=0.05$ and show the obtained spectral function and single-particle excitation minima in Fig.~\ref{fig:FSBD}(c)(d). The temperature is close to the SC transition temperature. In Fig.~\ref{fig:FSBD}(c) one can see a clear single-particle gap developing, which is minimal around the $X$ points. This developing gap is also reflected in the small value of $G(k,\tau=\beta/2)$ in Fig.~\ref{fig:FSBD}(d). Due to the tendency to form inter-layer spin-singlet pairs, the double occupancy is low within each orbital. In particular, from our simulations we find $\langle n_{x,\uparrow} n_{x,\downarrow}\rangle\approx0.038,\langle n_{z,\uparrow} n_{z,\downarrow}\rangle\approx0.13$. As a result, we expect the addition of a local Hubbard interaction in $d_{x^2-y^2}$ orbitals (which would introduce a sign problem) will not significantly change the phase diagram.

\par
\begin{figure}[htp!]
    \centering
    \includegraphics[width=\columnwidth]{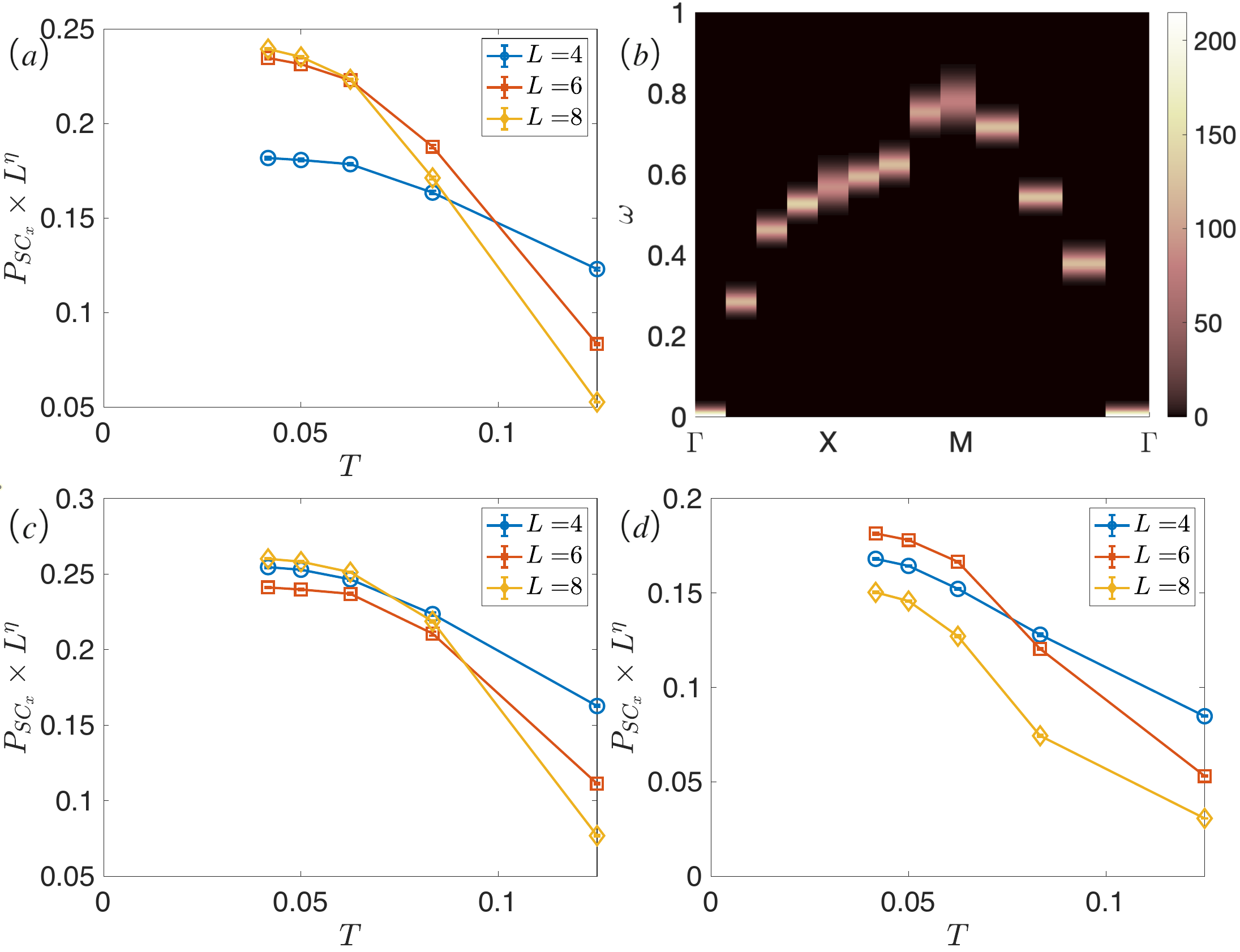}
    \caption{(a) Finite-size crossing of Cooper pair correlation in $d_{x^2-y^2}$ orbital indicates $T_c\lesssim 0.06$ for interacting Hamiltonian with parameters $t_{zz}^{\perp}=-0.635,J_{xz}=0.4,\mu=-0.2$. (b) Spectral function from SAC of imaginary time correlation function $C(\tau)=\langle \Delta^{\dagger}(\tau)\Delta(0)\rangle$ for $L=8,T=0.05$ case.
    (c)(d) The same plot as (a) with different chemical potential $\mu=0$ and $\mu=-0.4$ respectively.}
    \label{fig:SC}
\end{figure}

\noindent \textcolor{myblue}{\it{Superconductivity and the filling.-}}
To detect SC coming from the quarter-filled $d_{x^2-y^2}$ orbitals, we use the finite-size crossing of order parameter with critical exponent of entering Berezinskii–Kosterlitz–Thouless (BKT) phase $\eta=\frac{1}{4}$ to approximately identify the SC phase transition. The relevant pair correlation function is
\begin{eqnarray}
    &&P_{SC_x}=\frac{1}{L^4}\sum_{i,j}\langle \Delta^{\dagger}_{i,x}\Delta_{j,x}\rangle, \\
    &&\Delta^{\dagger}_{i,x}=c_{i,x}^{\dagger}l^{x}\sigma^{y}c_{i,x}^\dagger=c^{\dagger}_{i,1,\uparrow,x} c^{\dagger}_{i,2,\downarrow,x} - c^{\dagger}_{i,1,\downarrow,x} c^{\dagger}_{i,2,\uparrow,x}. \nonumber
\end{eqnarray}
When the system enters the SC phase from high temperature, $P_{SC_x}\times L^{\eta}$ is scale-invariant and should cross at the transition temperature $T_c$ for different $L$~\cite{zhang2022super,jiang2022monte,paiva2004critical}. From this finite-size crossing shown in Fig.~\ref{fig:SC}(a), we identify $T_c\lesssim 0.06$. 

In the SC phase, the $\Delta^{\dagger}$ excitation is gapless with a linear dispersion. In Fig.~\ref{fig:SC}(b) we show the spectral weight for the $\Delta^{\dagger}$ excitation, obtained from stochastic analytical continuation (SAC)~\cite{shao2023progress} of the imaginary-time correlation function $C(\tau)=\langle \Delta^{\dagger}(\tau)\Delta(0)\rangle$ at $T=0.05$. We can see a clear gapless branch emerging from $\Gamma$, whose high-energy part mixes with the single-particle excitations near the $X$ points (see Fig.~\ref{fig:FSBD}(c)). 

To study the filling dependence of SC, we change the chemical potential to $\mu =0$ (hole densities $\delta_{x^2-y^2}\approx0.33$, $\delta_{z^2}\approx0.012$) and $\mu=-0.4$ (hole densities $\delta_{x^2-y^2}\approx0.61$, $\delta_{z^2}\approx0.063$). The results are shown in Figs.~\ref{fig:SC}(c)(d). By comparing the tendency towards large-size limit at low temperature, one can conclude superconductivity prefers small-doping region, and a large doping in $d_{x^2-y^2}$ orbital will suppress and even eliminate the superconductivity.

\begin{figure}[htp!]
    \centering
    \includegraphics[width=\columnwidth]{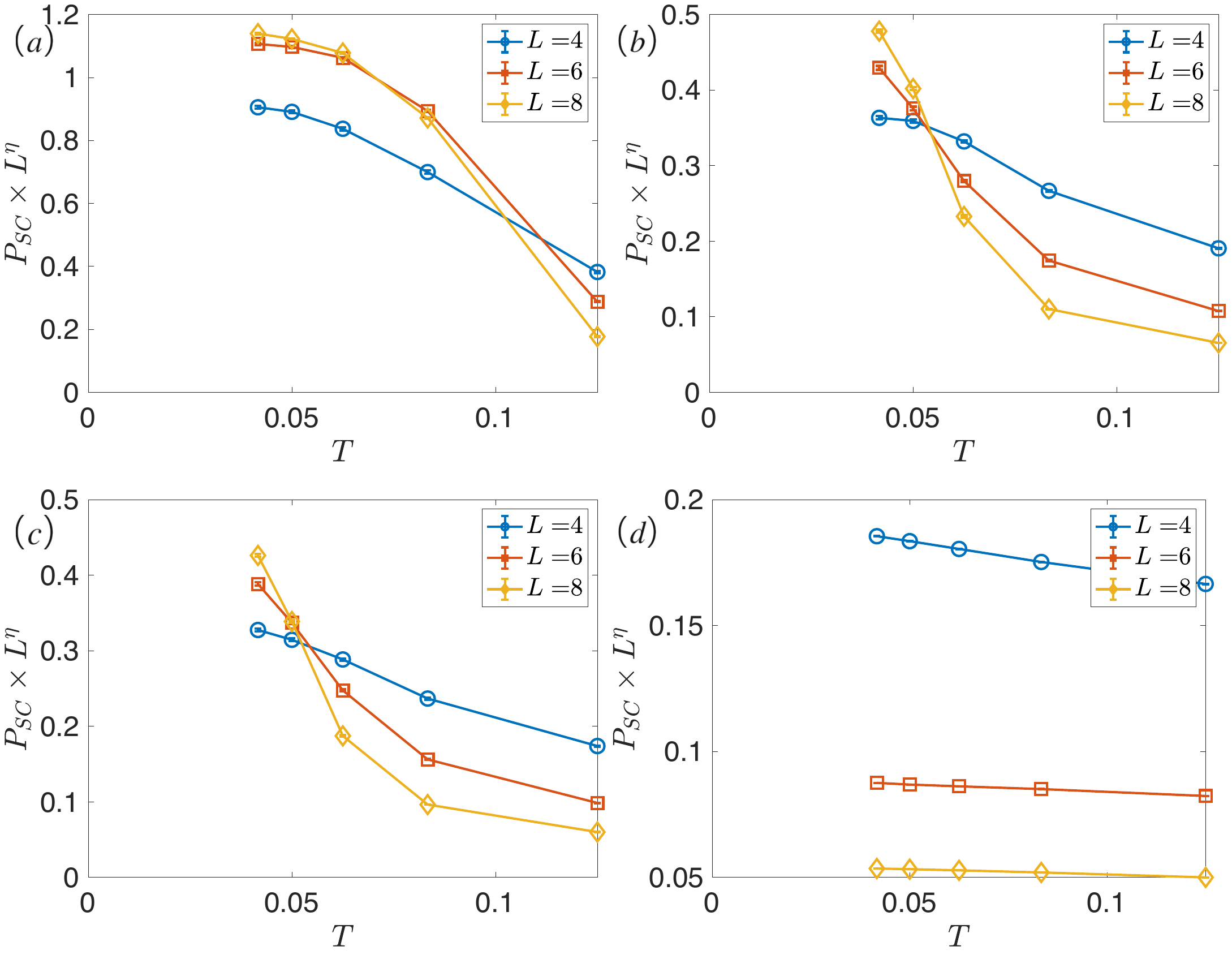}
    \caption{(a) Finite-size crossing of Cooper pair correlation indicates $T_c\lesssim 0.06$ for interacting Hamiltonian with parameters $t^\perp_{zz}=0,J_{xz}=0.4,\mu=-0.2$. (b)(c) Finite-size crossing with the same parameters as in (a), except for $J_{xz}=0.2$ in (b) and $J_{xz}=0.1$ in (c). (d) Absence of SC with $t^\perp_{zz}=-0.635,J_{xz}=0.2$.}
    \label{fig:SCnoh}
\end{figure}

\noindent \textcolor{myblue}{\it{Inter-layer tunneling and Hund's coupling.-}}
Next we investigate the role of the inter-layer tunneling $t^\perp_{zz}$ and Hund's coupling $J_{xz}$ in stabilizing SC. When setting the inter-layer tunneling to zero, $t^\perp_{zz}=0$, the hole density on the $d_{z^2}$ orbitals increases to $\delta_{z^2}\approx0.26$, while the $\delta_{x^2-y^2}\approx0.45$ remains approximately constant. We expect and observe SC comes from both types of orbitals, and hence the pair correlation function is $P_{SC}=\frac{1}{L^4}\sum_{i,j}\langle \Delta^{\dagger}_{i}\Delta_{j}\rangle, \Delta^{\dagger}_{i}=\sum_{m}\Delta^{\dagger}_{i,m}=\sum_{m}c^{\dagger}_{i,1,\uparrow,m} c^{\dagger}_{i,2,\downarrow,m} - c^{\dagger}_{i,1,\downarrow,m} c^{\dagger}_{i,2,\uparrow,m}$. Using identical parameters as in Figs.~\ref{fig:FSBD}(c)(d) and Figs.~\ref{fig:SC}(a)(b), except for $t^\perp_{zz} =0$, we obtain finite-size crossing as shown in Fig.~\ref{fig:SCnoh}(a) which indicates a similar SC transition temperature as in Fig.~\ref{fig:SC}(a). Keeping $t^\perp_{zz}=0$ and reducing the Hund's coupling strength to $J_{xz}=0.2,0.1$, we see from Figs.~\ref{fig:SC}(b)(c) that $T_c$ is largely unaffected, showing that the Hund's coupling is unimportant when pairing occurs in both the $d_{z^2}$ and $d_{x^2-y^2}$ orbitals. In contrast, when reinstating $t^\perp_{zz}$ to its DFT value used previously, we find that decreasing the Hund's coupling to $J_{xz}=0.2$ eliminates SC, as shown in Fig.~\ref{fig:SC}(d). Thus we can conclude with decreasing inter-layer tunneling $t^\perp_{zz}$ while keeping $J_{zz}$, the doping in $d_{z^2}$ orbital will reduce the dependency of SC on Hund's coupling $J_{xz}$ or artificially induced $J_{xx}$.

\noindent \textcolor{myblue}{\it{Inter-layer bond order and spin density wave.-}}We now study other candidate orders dicussed in the anti-unitary symmetry section before, and calculate structure factors for SC, BO and SDW with the same parameters as in Figs.~\ref{fig:FSBD}(c)(d) and Figs.~\ref{fig:SC}(a)(b). In particular, we calculate
\begin{eqnarray}
    &&S_{SC_x}(q)=\frac{1}{L^4}\sum_{j,k}e^{iq\cdot(r_j-r_k)} \langle \Delta^{\dagger}_{j,x}\Delta_{k,x}\rangle, \\
    &&S_{BO}(q)=\frac{1}{L^4}\sum_{j,k}e^{iq\cdot(r_j-r_k)} \langle \tilde{M}_{BO,j}\tilde{M}_{BO,k}\rangle, \nonumber\\
    &&S_{SDW}(q)=\frac{1}{L^4}\sum_{j,k}e^{iq\cdot(r_j-r_k)}  \langle M_{SDW,j}M_{SDW,k}\rangle\,, \nonumber
\end{eqnarray}
where $\tilde{M}_{BO,j}=\sum_{s,m}c_{j,1,s,m}^{\dagger}c_{j,2,s,m}+c_{j,2,s,m}^{\dagger}c_{j,1,s,m}-\left\langle c_{j,1,s,m}^{\dagger}c_{j,2,s,m}+c_{j,2,s,m}^{\dagger}c_{j,1,s,m} \right\rangle $, and $M_{SDW,j}=\sum_{l,m}(-1)^l (c_{j,l,\uparrow,m}^{\dagger}c_{j,l,\uparrow,m}-c_{j,l,\downarrow,m}^{\dagger}c_{j,l,\downarrow,m})$. 

\begin{figure}[htp!]
    \centering
    \includegraphics[width=\columnwidth]{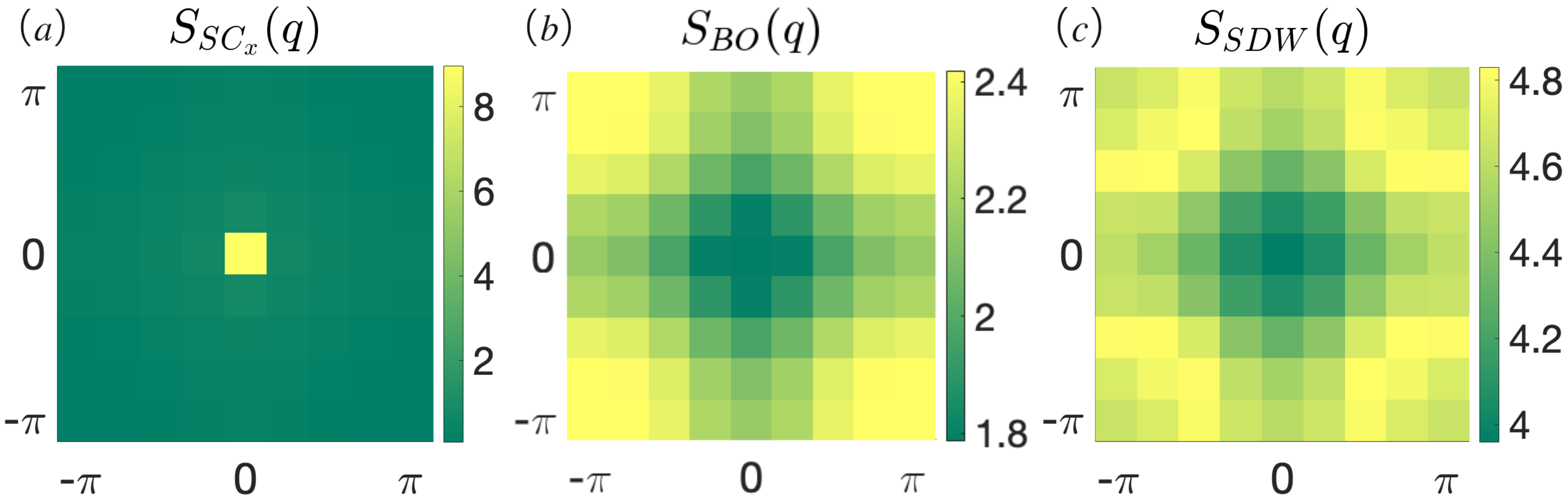}
    \caption{Static structure factors for SC (a), BO (b) and SDW (c) with the same parameters as in Fig.~\ref{fig:FSBD}(c)(d) and Fig.~\ref{fig:SC}(a)(b).}
    \label{fig:Sq}
\end{figure}

The static structure factors are shown in Fig.~\ref{fig:Sq}. In Fig.~\ref{fig:Sq}(b) we see that the BO structure factor is maximal at $(\pi,\pi)$, in agreement with our analysis of the dominant ordering instabilities based on the Kramers anti-unitary symmetries in the strong-interaction limit. Besides, we also observe a peak in the static spin structure factor, indicating fluctuating SDW order at a momentum close to $(\frac{\pi}{2},\frac{\pi}{2})$. Since we decouple the interaction in spin channel, the SDW order here corresponds to the auxiliary-field saddle point and the momentum peak is determined by the details of integrating out fermions.

\noindent \textcolor{myblue}{\it{Conclusion and discussion.-}}
Different from the $t$-$J$-$J_{\perp}$ type of approximation like in Refs.~\cite{qu2024bilayer,xue2024magnetism}, we take another strong-coupling approximation by first combining the dominant hopping $t^\perp_{zz}$ and onsite repulsion $U_{zz}$ to form $J_{zz}$ interaction, and seeing the left parts of Hamiltonian related with $d_{z^2}$ orbitals as higher-order contributions. To reach the SC phase in our simulation, we use the parameter $J_{zz}=0.4$ and tune $J_{xz}$ from $0.4$ to $0.1$ at two limits $t_{zz}^{\perp}=-0.635$ and $t_{zz}^{\perp}=0$. The numerical results are concluded below:
\begin{enumerate}
\item  With the $t^\perp_{zz}$ and doping from DFT~\cite{luo2023bilayer}, we find a SC phase with transition temperature $T_c\lesssim 0.06$ coming from the $d_{x^2-y^2}$ orbital, generated by a simplified interaction with $J_{zz}=J_{xz}=0.4$ together with an artificial $J_{xx}=0.4$. A larger hole-doping or smaller $J_{xz}$ eliminates superconductivity.
\item Taking inter-layer tunneling $t^\perp_{zz}=0$ limit increases the hole density in the $d_{z^2}$ orbital. And a SC phase with similar transition temperature $T_c\lesssim 0.06$ emerges from both $d_{z^2}$ and $d_{x^2-y^2}$ orbitals, where $T_c$ is insensitive to a decrease in the Hund's coupling strength $J_{xz}$.
\end{enumerate}
We highlight the phase diagram study of the sign-problem-free model can be understood qualitatively by analyzing the (approximate) Kramers anti-unitary symmetries of the auxiliary-field-decoupled Hamiltonian~\cite{zhang2025constraints}, which implies that (1) the leading SC instability is in the spin-singlet layer-triplet channel, and (2) there is a competing BO at momentum $(\pi,\pi)$. The fluctuating BO is accompanied by fluctuating SDW order near momentum $(\frac{\pi}{2},\frac{\pi}{2})$, which is determined by the details of integrating out fermions.

The BO instability at $(\pi,\pi)$ found in this work is a promising candidate for the unknown density wave observed in experiment~\cite{zhao2025pressure,khasanov2025pressure}.
The findings above also suggest two avenues for enhancing superconductivity driven by an antiferromagnetic interaction between two subspaces (e.g., layer, valley,...): (1) strongly breaking particle-hole symmetry, and (2) reducing the inter-subspace tunneling (to avoid an instability towards BO). The latter is closely related to the important role of the apical oxygen vacancies in stabilizing superconductivity in the bilayer nickelates~\cite{dong2024visualization}, as slight vacancies reduce the inter-layer tunneling and the Kramers anti-unitary symmetry for SC is insensitive with such a translational symmetry breaking.  More generally, since an antiferromagnetic super-exchange interaction closely depends on tunneling, looking for a material where pairing relies on a tunneling-independent antiferromagnetic interaction (e.g. RKKY interaction) could be a promising direction to find new high-$T_c$ superconductors.

\begin{acknowledgements}
{\noindent \it Acknowledgements.---}
I thank Nick Bultinck for an inspiring discussion on nickelates superconductor. I thank Nick Bultinck for careful proofreading, Jiangping Hu for comments and Zi Yang Meng, Kai Sun for useful suggestions on the draft.
This research was supported by the European Research
Council under the European Union Horizon 2020 Research and Innovation Programme via Grant Agreement No. 101076597-SIESS. The computational resources and services used in this work were provided by the VSC (Flemish Supercomputer Center), funded by the Research Foundation - Flanders (FWO) and the Flemish Government. I acknowledge EuroHPC Joint Undertaking for awarding me access to MareNostrum5 hosted by Barcelona Supercomputing Center, Spain.
\end{acknowledgements}

\bibliography{Nick_QMC}

\end{document}